\documentclass[preprintnumbers,amsmath,amssymb]{revtex4}
\usepackage{graphicx}
\begin{document}
\preprint{}
\title{de Sitter Black Holes with Either of the Two Horizons as a Boundary}
\author{Andr\'es Gomberoff}
\affiliation{Centro de Estudios Cient\'{\i}ficos (CECS), Valdivia, Chile.}
\author{Claudio Teitelboim}
\affiliation{Centro de Estudios Cient\'{\i}ficos (CECS), Valdivia, Chile.}

\date{February 25, 2003}

\begin{abstract}
The action and the thermodynamics of a rotating black hole
in the presence of a positive cosmological constant are analyzed.
Since there is no spatial infinity, one must bring in, instead, a platform where
the parameters characterizing the thermodynamic ensemble
are specified. In the present treatment the platform in question is taken to be
one of the two horizons, which is considered as a boundary.
If the boundary is taken to be the cosmological horizon one deals with
the action and thermodynamics of the black hole horizon.
Conversely, if one takes the black hole horizon as the boundary, one deals with
the action and thermodynamics of the cosmological horizon. The two systems are different.
 Their energy and angular momenta are equal in magnitude but have opposite sign.
In either case, the energy and the angular momentum are obtained as surface terms
on the boundary, according to the standard Hamiltonian procedure. The temperature
and the rotational chemical potential are also expressed in terms of magnitudes on the boundary.
If, in the resulting expressions, one continues the cosmological constant to negative values, the black hole
thermodynamic parameters defined on the cosmological horizon coincide with those calculated at spatial infinity in the
asymptotically anti-de Sitter case.

\end{abstract}

\maketitle

\section{introduction}

The reported observational evidence for a positive cosmological constant\cite{obs} has led to
renewed interest in de Sitter space \cite{Witten, Strominger, Balasubramanian, 
Banks, Goheer, Bousso, Shankaranarayanan}. 
It is natural in this
context to analyze the thermodynamics of de Sitter space in the presence of
a black hole. It has been known for a long time  that
if one  uses the Euclidean Schwarzschild--de Sitter solution (or, more generally,
the Euclidean Kerr--de Sitter solution) to
provide thermodynamical information, one finds that the time periods
required to avoid a conical singularity, at both  the cosmological and black
hole horizons, do not match. This is physically interpreted as indicating
that the two horizons are not in thermal equilibrium and that, for example,
they both emit Hawking radiation at the corresponding temperatures.
An observer somewhere in space would then
receive a beam of radiation coming from the black hole and,
at the same time,  radiation coming from the
cosmological horizon would arrive to him from all directions\cite{GH}.

From the point of view of the action principle, the fact that the
Schwarzschild--de Sitter solution has a conical
singularity means that the empty space field equations are not satisfied
everywhere. If one arranges the period of the time variable so as to have no
conical singularity at the cosmological horizon, the field equations will be
satisfied there but will not be satisfied at the black hole horizon.
Conversely, if the role of the horizons is interchanged, the field equations
will not be satisfied at the cosmological horizon.

The main purpose of this article is to point out that this apparent
difficulty is rather a virtue, and, at  that,  one which was to be expected
from the point of view of the action principle and thermodynamics. Indeed, if one uses an
extremum of the action to evaluate the path integral in the semiclassical
approximation, one needs to hold fixed those variables which will become the
argument of the partition function once it is evaluated. By the very meaning
of ``holding fixed", those variables are not varied in the action
principle. Thus, for example, for a black hole in asymptotically flat space,
one may hold fixed the $1/r$ part of the components of the metric
which are determined by the mass. Then, one is dealing with the
microcanonical ensemble, where the partition function depends on the total
energy. It would be wrong to demand that the partition function thus
obtained should have an extremum with respect to variations in the $1/r$
piece of the spatial metric, because then one would obtain a particular
value for the mass, i. e., $m=0$ (and infinite temperature), and thus would not have enough information
to develop the thermodynamics of the system.

For the case at hand, there is no notion of spacelike infinity, but the
problem itself indicates what to do. One may fix appropriate components of
the metric at either the cosmological or black hole horizons. If one chooses
the cosmological horizon as the place where the variables are held fixed,
there will be no field equations to satisfy there.
Then the cosmological horizon will be the analog of
spatial infinity in the asymptotically flat case, where the ``observer" sits
(or, more precisely, the analog of a very large sphere whose radius is
eventually allowed to grow without limit). The problem one is solving then
will be the thermodynamics of a black hole contained in a space of a given
cosmological radius (``box", ``boundary"). Conversely, if the variables are fixed
at the black hole horizon, it is then that horizon which acts as the
boundary. One would then be discussing the thermodynamics of a cosmological
horizon with the black hole acting as the boundary. Changes in the black
hole variables would then not be subject  to dynamics but rather would
correspond to changes that the ``external observer" decides to make in the
environment.

This discussion, which was first given in \cite{CT},
 shows that one should be able to use the Schwarzschild--de Sitter solution
(and also, the Kerr--de Sitter solution)  as a true extremum of two different action
principles which correspond to two different thermodynamical problems. One problem is the
thermodynamics of a black hole horizon with a cosmological boundary. The
other is the thermodynamics of a cosmological horizon with a black hole
boundary. It turns out, as we shall see below, that the physical
properties of the two systems have striking differences.

 The paper is organized as follows. Section II, which is the bulk of the article, contains the analysis of the
Hamiltonian action principle and the identification and evaluation of the energy and the
angular momentum, as well as the entropy, as surface terms for a black hole solution in the presence of a positive
cosmological constant.  The energy $U$ and angular momentum $J$ of Kerr--de Sitter space
are evaluated as surface terms at each of the two horizons. The sign of both $U$ and $J$
is reversed if the role of the horizons is interchanged. Interestingly enough, the expressions
for the energy $U_+$ and angular momentum $J_+$, which are surface terms on the cosmological
horizon, turn out to be identical to the surface integrals at spacelike infinity for the
Kerr--{\it anti}--de Sitter solution, if, in those expressions, one replaces the cosmological
constant by its negative, or, what is the same, if one continues the anti--de Sitter
radius to imaginary values. Next, section III contains a brief discussion of the essential thermodynamical
aspects of the system. Finally, the properties of the Kerr--de Sitter geometry that are used in the main text are
summarized in Appendix A, while Appendix B contains an alternative treatment
of the spherically symmetric case (no rotation).

\section{Action Principle}
\label{one}

Since our main interest is gravitational thermodynamics, we shall consider Euclidean line elements, $ds^2$, and the Euclidean action, $I$, throughout. It is, however,
important to emphasize that the identification of the energy and the angular momentum as surface terms could be achieved in the same way with Lorentzian signature. It
is only for the entropy that it is necessary to work with  Euclidean signature. Our sign convention for the Euclidean action is that, in the microcanonical ensemble, the
semiclassical approximation for the entropy $S$, coincides with the value of $I$ on--shell. The Euclidean black hole spacetimes have the topology of $S_2\times
S_2$. The first $S_2$ may be taken to be the ordinary spatial $2$--sphere with coordinates $\theta$ and $\phi$. The second $S_2$ will be spanned by the
coordinates $r$ and $t$. The radial coordinate $r$ will run between the black hole and cosmological horizons, $r_+$, $r_{++}$, which correspond to both
geographical poles of the sphere. The coordinate $t$ plays the role of the azimuthal angle. When one horizon is treated as a boundary, that point is
removed from the sphere, which becomes a disk. No equations of motion are imposed on the boundary, since the fields are not varied but held fixed. On the other
horizon, which is not treated as a boundary, the fields are varied in the action principle and, therefore, the action principle must be such that the Einstein equations hold
there.

\subsection*{Boundary terms}

We will be specially interested in the surface terms at the horizons which arise in the variation of the action. We will assume that near each of the horizons the metric
may be written in the form
\begin{equation}
\label{metric}
ds^2 \approx (N^{\perp}_{,\rho})^2 \rho^2 dt^2 + d\rho^2 + g_{\theta\theta}\ d\theta^2
+ g_{\phi\phi}(d\phi+N^{\phi}dt)^2 \ .
\end{equation}
In this local coordinate patch, the corresponding horizon is located at $\rho=0$, where
$N^{\perp}=0$, and the coefficients $N^{\perp}_{,\rho}$ and $N^{\phi}$ at $\rho=0$ do not depend on $t$, $\theta$ and $\phi$, whereas $g_{\theta\theta}$,
and $g_{\phi\phi}$, may depend on $\theta$ at $\rho=0$. As ones moves away from the horizon $\rho$ increases and so does $N^{\perp}$. Thus
$N^{\perp}_{,\rho}$ at $\rho=0$, appearing in (\ref{metric}), is positive.
As discussed in Appendix A, this form of the line element near the horizon includes the Kerr--de Sitter solution and therefore it suffices for the purpose of this paper. In particular, the independence of the metric components at $\rho=0$ on $t$ and $\phi$ is a remnant, imposed as a boundary condition, of the fact that the Kerr solution is stationary and axisymmetric. The off--shell metric away from the horizons is arbitrary.

As shown by Eq. (\ref{metric}) the horizon is at $\rho=0$, which is the origin
of a polar coordinate system in the $(\rho,t)$--plane. Near $\rho=0$, the $t=$constant surface rotates around the horizon during the time interval $t_2-t_1$ by a
``proper angle"
\begin{equation}
\Theta= N^{\perp}_{,\rho} (t_2-t_1) \ .
\label{angle}
\end{equation}

Now, consider the Hamiltonian action,
\begin{equation}
I_{\mbox{{\tiny HAMILTONIAN}}}=\int d^4 x \left( \pi^{ij}\dot{g}_{ij} - N^{\perp}{\cal H}_{\perp} - N^{i}{\cal
H}_{i} \right) \ .
\label{hamaction}
\end{equation}
If one varies $I_{\mbox{{\tiny HAMILTONIAN}}}$ one finds both ``bulk terms" (integrals over space and time), and surface terms,
at each of the horizons. The bulk terms vanish when the Einstein equations hold, and the surface terms take the form\cite{BTZ},
\begin{equation}
-4\pi(t_2-t_1)\left.\int d\theta\left[ \ N^{\perp}_{,\rho} \delta g^{1/2}
 \ + N^{\phi}\delta\pi_{\phi}^{\ \rho} \right]\right|_{\rho=0} \ ,
\label{surface}
\end{equation}
where $g=g_{\theta\theta}g_{\phi\phi}$ is the determinant of the angular metric.
As we now pass to discuss, the surface term (\ref{surface}) is handled differently depending on whether the horizon at which it is evaluated is treated as dynamical and
therefore endowed with thermodynamical properties or as a boundary, where some fields are prescribed.

\subsection*{Horizon dynamical: surface fields not fixed. Entropy.}

If the horizon is treated as dynamical, one does not hold fixed anything at $\rho=0$, and therefore the action should have an extremum with respect to variations of
$g^{1/2}$ and $\pi^{\phi}_{\rho}$. The extremization of the hamiltonian action (\ref{hamaction}) with respect to $g^{1/2}$ would yield the undesirable result
$\Theta=0$ for the opening angle. One wishes instead to have an action principle that gives
\begin{equation}
\Theta=2\pi \ ,
\label{noconical}
\end{equation}
which means that there is no conical singularity and therefore, that, on--shell,
the manifold is smooth at $\rho=0$, as it is at any other point.
This is accomplished by adding to $I_{\mbox{{\tiny HAMILTONIAN}}}$ the surface term\cite{BTZ},
\begin{equation}
4\pi A = 8\pi^2\int d\theta g^{1/2} \ ,
\label{add}
\end{equation}
which is, of course, the horizon entropy,
 as will be recalled below (we use units in which $16\pi G =1$, so that the $4\pi$ appearing in (\ref{add}), may also be written as $1/4G$).

It is not necessary to add any additional term to properly account for the extremization
with respect to $\pi_{\ \phi}^{\rho}$. Indeed, one obtains from (\ref{surface})
\begin{equation}
N^{\phi} = 0 \ ,
\label{zeronphi}
\end{equation}
which states that the angular coordinate system  should be taken as ``co--rotating"
with the dynamical horizon, as can always be done.  Note that the metric given in Appendix A does not obey (\ref{zeronphi}). To that end,
one would need to make the change of coordinates,
\begin{equation}
\phi =  \phi'-N^{\phi}_{\mbox{\tiny horizon}}t \ .
\label{ccc}
\end{equation}

\subsection*{Horizon as boundary: surface fields fixed. Energy and angular momentum}

In order to avoid unnecessary cluttering of the equations, we shall take
the black hole horizon at $r_+$ as the dynamical one and the cosmological horizon at
$r_{++}$ as the boundary. After the analysis is carried out it will be indicated
how the conclusions are changed if the role of the horizons is reversed. Those changes
will be simple but significant.

\subsubsection*{Angular momentum}
When the surface fields are fixed, the surface term itself indicates which
variables should be held fixed, and, also, gives their physical meaning.
The piece of the surface term which contains $N^{\phi}$ lends itself to immediate
treatment. Indeed, that surface term comes from ${\cal H}_\phi$, which is the generator
of purely spatial deformations of the hypersurface $t=$constant along the direction $\phi$. If
$N^{\phi}$ is a constant over the boundary, as it happens in the present case, the deformation at the boundary
over the interval $(t_2-t_1)$ is a spatial rotation by an angle
$N^{\phi}(t_2-t_1)$.
The coefficient is, therefore, the variation of the conjugate ``global charge", which is the angular momentum $J$.
Thus, we find,
\begin{equation}
J_+ = -\int_{\rho=0} d\theta \pi^{\rho}_{\phi} =
\int_{r_{++}} d\theta \pi^{r}_{\phi} \ ,
\label{j}
\end{equation}
for the angular momentum of the horizon at $r_+$ evaluated as a surface integral
at the boundary $r_{++}$. The change in sign when passing from
the coordinate $\rho$ to $r$ arises because $ \pi^{i}_{j}$ is a tensor
density and $dr/d\rho$ is negative at $r_{++}$.
When expressed in terms of the Kerr--de Sitter metric
parameters $m$, $a$ (see Appendix A), Eq.(\ref{j}) gives
\begin{equation}
J_+ = \frac{ma}{Z^2} \ .
\label{jkerr}
\end{equation}

\subsubsection*{Energy}

Next we turn to the surface term
\begin{equation}
-4\pi(t_2-t_1)\int d\theta \ N^{\perp}_{,\rho} \delta g^{1/2} \ ,
\label{energy}
\end{equation}
which comes from the variation of ${\cal H}_{\perp}$. Normally, {\it i.e.},
in the asymptotically flat or anti--de Sitter cases, one ``pulls the delta through to the left" to achieve the form
\begin{equation}
(t_2-t_1) \delta U \ .
\label{e}
\end{equation}
Then $U$ is identified as the total energy because it is conjugate
to the time $t$. [We have used the letter $U$ because it is the customary denomination
of the internal energy in thermodynamics. As we will see, $U$ will, in general,
not coincide with the mass parameter $m$ of the Kerr--de Sitter, or even
Schwarzschild--de Sitter solutions].

In the present case, it will turn out that one can pull the delta through only if one adds
to the surface term (\ref{energy}), coming from ${\cal H}_{\perp}$ an additional contribution coming from ${\cal H}_{\phi}$. This means that, on the boundary, what one would like to call a ``pure time displacement" is not perpendicular to the $t=$constant surface, but includes, in addition, a specific spatial rotation. As it will be explained further below, this fact reflects  a noteworthy connection between the de Sitter and anti--de Sitter cases.

The analysis proceeds as follows. From the expressions given in Appendix A, one can write
the surface term (\ref{energy}) as
\begin{equation}
4\pi(t_2-t_1)\left\{ \frac{H'(r_{++})}{(r_{++}^2+a^2)}\ \delta
\left[\frac{(r_{++}^2+a^2)}{Z}\right] \right\}
\ . \label{a}
\end{equation}
To pull the delta through to the left we observe the identity,
\begin{equation}
4\pi\frac{H'(r_{++})}{(r_{++}^2+a^2)} \ \delta \left[\frac{(r_{++}^2+a^2)}{Z}\right] =
\delta\left[ \frac{m}{Z^2}\right] - \left( \frac{aZ}{r_{++}^2+a^2} - \frac{a}{l^2} \right) \ \delta \left[\frac{ma}{Z^2}\right] \ ,
\label{identity}
\end{equation}
(which also holds if $r_{++}$ is replaced by $r_+$ throughout)
and, hence, we find that the surface term coming from ${\cal H}_{\perp}$ (\ref{energy})
may be written in the form,
\begin{equation}
(t_2-t_1) \left[\delta U - \left(\frac{aZ}{r_{++}^2+a^2} - \frac{a}{l^2}  \right) \delta J \right] \ ,
\label{s}
\end{equation}
with
\begin{equation}
U = \frac{m}{Z^2} \ .
\label{U}
\end{equation}

Since in arriving at expressions for the total energy (\ref{U}) and angular momentum (\ref{jkerr}) we have only used the form of the
Kerr--de Sitter metric near the boundary at $r_{++}$, these formulas should also hold for any configuration that approaches the
Kerr-de Sitter metric at $r_{++}$. Thus, expressions (\ref{U}) and (\ref{jkerr}) have the same content as the standard ADM surface integrals
for asymptotically flat space (see, for example \cite{RT}) or their generalization to asymptotically anti--de Sitter space\cite{HT}. A closer
formal analog, but with the same physical content,  of the ADM type expressions would be obtained by writing the surface terms for
$J$ and $M$ directly in terms of the spatial metric and its conjugate momentum. This has been already achieved for $J$ in expression (\ref{j}).
The corresponding expression  for $U$ can be developed easily for the case
of spherical symmetry, and it is given in Appendix B. Straightforward efforts to do the same in the case of rotation did not prove successful.

\subsubsection*{Chemical potential. Relationship with anti--de Sitter space}

As mentioned before, the presence of the $\delta J$ term in (\ref{s}), means, geometrically,
that the ``global time displacement" generated by the total energy $U$ contains a spatial rotation as well as a deformation
of the constant time surface normal to itself by an angle
\begin{equation}
 (t_2-t_1) \left(\frac{aZ}{r_{++}^2+a^2} - \frac{a}{l^2}  \right) \ ,
\label{a1}
\end{equation}
which may be rewritten in terms of the shift $N^{\phi}(r_{++})$ of the Kerr--de Sitter metric,
given in Appendix A, as
\begin{equation}
-(t_2-t_1) \left(N^{\phi}(r_{++}) + \frac{a}{l^2}\right)
\label{a2}
\end{equation}
One may insert expression (\ref{a1}) in (\ref{s}) to write the total surface term (\ref{surface}) in the form
\begin{equation}
(t_2-t_1)(\delta U - \Omega \delta J) \ ,
\label{surface2}
\end{equation}
with,
\begin{equation}
\Omega = N^{\phi}(r_+) - \frac{a}{l^2} \ .
\label{omega}
\end{equation}
>From the point of view of thermodynamics, or, of  classical mechanics if one wants, the presence of the extra rotation has the consequence that the conjugate $\Omega$ to $J$ is not the relative angular velocity
\begin{equation}
N^{\phi}(r_{+}) - N^{\phi}(r_{++}) \ ,
\label{dif}
\end{equation}
of the rotating black hole horizon $r_+$ relative to the cosmological horizon $r_{++}$, as one might have naively expected.
Instead, one may describe the chemical potential $\Omega$ given by (\ref{omega}) as the angular velocity of the black hole horizon relative to
$r=\infty$, because the term $-a/l^2$ is what one obtains if one sets $r=\infty$ in $N^\phi$ of the Kerr--de Sitter metric given in (\ref{kds}). Of course
$r=\infty$ is not in the Euclidean section, but one may give meaning to this formal substitution by saying that expression (\ref{omega}) for $\Omega$, and the expressions (\ref{U}), (\ref{jkerr}) for $U$ and $J$ are precisely the analytical continuations of the expressions for the anti--de Sitter case obtained in Ref. \cite{HT}, if one takes the anti--de Sitter radius to be imaginary.
It is remarkable that the present treatment, based on identifying the energy and the angular momentum of the black hole horizon as surface terms on the cosmological
horizon should give precisely
the analytic continuation of the anti--de Sitter expressions with the generators normalized
according to the standard $O(4,1)$ Lie algebra structure constants.
[Throughout the discussion of the energy, angular momentum and chemical potential, we  use the angular coordinate employed in Appendix A, which does not fulfill (\ref{zeronphi}). In this way, the contrast between the actual result (\ref{omega}) and the difference of shifts given by (\ref{dif}), whose value
is invariant under the coordinate change (\ref{ccc}), is brought out more clearly. In any case, all the surface terms occur at the boundary, which is at $r_{++}$
in this case].

\subsection*{Reversing the roles of the horizons}

We now consider the case in which the black hole horizon $r_+$ is taken as the boundary
and the cosmological horizon $r_{++}$ is taken as dynamical. The analysis goes through following the exact same reasoning and steps as before. One finds for the
energy $U_{++}$ and angular
momentum $J_{++}$
\begin{eqnarray}
U_{++} &=& -\frac{m}{Z^2} = -U_+ \label{U++} \\
J_{++} &=& -\frac{ma}{Z^2} = -J_+ \label{J++} \ .
\end{eqnarray}
The sign changes arise ultimately because when the Hamiltonian action is written
as an integral over a globally defined radial variable $r$, the surface term which
appears in the variation is a difference of two integrals of the same form, one at $r_{++}$, the
other at $r_+$. Technically, in the present setting, where for the sake of
geometrical clarity we have introduced a local variable $\rho$, which increases from zero to
a positive value as one moves away from each horizon, the changes in sign arise because
$dr/d\rho$ is positive at $r_+$ and it is negative at $r_{++}$. In the case of the angular
momentum this was already mentioned in Eq. (\ref{j}). For the energy $U$, the change of sign
follows from Eq. (\ref{a11}) of Appendix A.

The chemical potential (angular velocity) conjugate to $J_{++}$, $\Omega_{++}$,
is now given by
\begin{equation}
\Omega_{++} = N^{\phi}(r_{++}) - \frac{a}{l^2} \ .
\label{omega++}
\end{equation}
Again, this $\Omega$ is not the angular velocity of the cosmological horizon
relative to the black hole horizon. The strict analogy to what was proposed
for the black hole horizon would be to say that $\Omega_{++}$ is an angular velocity
relative to something beyond, {\it i.e.}, inside the black hole horizon. This, indeed, may be said,
because it so happens that $- \frac{a}{l^2}$ also equals $N^{\phi}$ at $r=0$.
However, in this case, there appears to be no obvious interpretation in terms of an analytic continuation in the de Sitter radius $l$.
Also, the surface $r=0$ is timelike in the Lorentzian section, but, at the moment of this writing, we
do not know either, if, and how, the results of this article are related to discussions
in terms of asymptotia in time (see, for example, \cite{Witten, Strominger, Balasubramanian, Bousso}).

Quite independently of the above issues, which manifest themselves in the presence
of rotation, we should emphasize that the consideration of a black hole in de Sitter
space leads naturally, as a particular case, to define the energy and angular momentum
of de Sitter space by removing the line $r=0$, corresponding to the worldline of a geodesic observer.
Both $U$ and $J$ vanish in this case. One may apply to de Sitter space with the line
$r=0$ removed a general de Sitter transformation. One would obtain then, again,
a de Sitter space, but with the line $r=0$ of another de Sitter coordinate system removed.
This would be a different space which would also have $U=J=0$. In that sense
one might say that the ``vacuum" is degenerate. Again,  it is beyond the scope of this paper to
attempt to relate these elementary comments to the ample existing literature on de Sitter
vacua\cite{vac, Bousso}.

\section{Thermodynamics}

If one fixes the energy and the angular momentum, one is dealing with the microcanonical
ensemble in thermodynamics. In the semiclassical approximation, the value of the
action in that ensemble gives the entropy. An important advantage of using the Hamiltonian action,
as we have done in this paper, is that the bulk term given by Eq. (\ref{hamaction})
vanishes on--shell for a stationary solution, such as the Kerr-de Sitter metric.
Therefore, one finds from (\ref{add}) the expected result,
\begin{equation}
S=4\pi A = \frac{1}{4G} A \ ,
\label{entropy}
\end{equation}
for the entropy $S$.
One may also use the canonical ensemble, in which case one adds to the action the term
$-\beta U$, where the inverse temperature $\beta$ is now held fixed. If one demands this
new action (which is the negative of the Helmholtz free energy) to have an extremum under
variations of $U$, one finds the standard thermodynamical result
\begin{equation}
\beta=\left.\frac{\partial S}{\partial U}\right|_J \ ,
\label{beta}
\end{equation}
which combined with Eq. (\ref{s}) gives
\begin{equation}
\beta=t_2-t_1  \ ,
\label{betat}
\end{equation}
and therefore fixes the time period. It is important to
clarify here the geometrical meaning of the time $t$. Indeed, in asymptotically
flat space one is satisfied with the relation (\ref{betat}) between the temperature and the
Euclidean time $t$, because one has assumed that the lapse function $N^\perp$ tends to
unity at infinity. Here, Eq. (\ref{betat}) holds with the standard form of the Kerr--de Sitter metric,
written in Appendix A, for which the rescaled lapse function,
\begin{equation}
N=f^{-1} N^{\perp} \ ,
\label{resclapse}
\end{equation}
introduced in Appendix B, is set equal to unity at the horizon. A similar rescaling
of the lapse function comes in anti--de Sitter space. In all three cases, the time $t$
is the ``Killing time". In asymptotically flat space $t$ coincides with the proper time displacement but
in the de Sitter case it does not coincide with the proper angle $\Theta$ introduced in Eq. (\ref{angle}),
 and, in anti--de Sitter,  it does not coincide with the hyperbolic analog of $\Theta$.

\acknowledgments

CECS is a Millennium Science Institute. It is also funded in
part by grants from Fundaci\'on Andes and the Tinker Foundation.
Support from Empresas CMPC is gratefully acknowledged as well.
AG gratefully  acknowledges support from FONDECYT grant
1010449  and individual support from Fundaci\'on Andes.
AG and CT  acknowledge partial funding
under FONDECYT grant 1010446.

\appendix
\section{The Kerr--de Sitter geometry}
\label{aa}
The Kerr--de Sitter line element may be written as
\begin{equation}
\label{kds}
ds^2 = N^2 dt^2 + \frac{R^2}{H}dr^2 + \frac{R^2}{\Delta}d\theta^2
+ g_{\phi\phi}(d\phi+N^{\phi}dt)^2 \ ,
\end{equation}
with,
\begin{eqnarray}
\label{def}
N^2 &=& \frac{R^2 H \Delta}{\Delta(r^2+a^2)^2-Ha^2\sin^2\theta} \ ,  \\
N^\phi &=& a Z \frac{H-\Delta(r^2+a^2)}{\Delta(r^2+a^2)^2-Ha^2\sin^2\theta} \label{nphi}\ , \\
g_{\phi\phi} &=& \frac{\sin^2\theta}{R^2Z^2}\left[
\Delta(r^2+a^2)^2-Ha^2\sin^2\theta\right] \ ,
\end{eqnarray}
where
\begin{eqnarray}
H &=& (r^2+a^2)\left(1-\frac{r^2}{l^2}\right) - \frac{m}{8\pi} r \ , \\
\Delta &=& 1+\frac{a^2}{l^2}\cos^2\theta \ , \\
R^2 &=& r^2 + a^2\cos^2\theta \ , \\
Z &=& 1+\frac{a^2}{l^2} \ .
\end{eqnarray}
The horizon radii $r_+$, $r_{++}$, are functions of $m$ and $a$ which solve,
\begin{equation}
H(r) = 0 \ .
\label{hor}
\end{equation}
Now consider the following change of coordinates,
\begin{equation}
\rho = \pm\frac{R(r_h)}{r_h}\int_{r_h}^{r}\frac{\tilde{r}d\tilde{r}}{\sqrt{H(\tilde{r})}} \ ,
\label{change}
\end{equation}
where $r_h$ is equal to either $r_+$ or $r_{++}$. The sign must be taken
to be positive when $r_h=r_+$ and negative when $r_h=r_{++}$.
As $\rho\rightarrow 0$, to leading order in
$\rho$, the metric takes the form (\ref{metric}), where
\begin{equation}
N^{\perp}_{,\rho} = \pm\frac{1}{2(r_h^2+a^2)} \left.\frac{dH}{dr}\right|_{r_h} \ \ .
\label{a11}
\end{equation}
Here the sign follows the same prescription as above.
Expression (\ref{a11}) was used in the derivation of Eq. (\ref{a}) in the main  text.

\section{Spherical case revisited}

 We assume that  the metric  becomes spherically symmetric as it approaches the boundary,
\begin{equation}
ds^2= N^2 (\gamma_{,\rho})^2 dt^2 + d\rho^2 + \gamma^2 d\Omega^2 \ \ \ , \label{boundaryform}
\end{equation}
where $d\Omega^2$ is the metric of the 2-sphere. The rescaled lapse $N$ and the coefficient
$\gamma$ are functions of $\rho$ only. The boundary is a horizon, whose radius $\rho=\rho_h$ is defined by,
\begin{equation}
\left.\gamma^2_{,\rho} \right|_{\rho_h} = 0 \ ,
\label{horizon}
\end{equation}
which means that at a horizon the area of the $2$--sphere is extremal.
The lapse function, $N^{\perp}= N |\gamma_{,\rho}|$ vanishes on the horizon.
If we redefine the hamiltonian generator,
\begin{equation}
\tilde{{\cal H}}_{\perp} = {\cal H}_{\perp}   |\gamma_{,\rho}|  \ ,
\label{ham}
\end{equation}
the associated Lagrange multiplier is $N$, so that the
corresponding term in the hamiltonian is
\begin{equation}
\int d^3 x(N^{\perp}{\cal H}_{\perp}) =  \int d^3 x(N{\tilde{\cal H}}_{\perp}) \ ,
\label{action}
\end{equation}
where
\begin{equation}
\tilde{{\cal H}}_\perp
= 2\sin\theta\   |\gamma_{,\rho}|  \left(2\gamma\gamma_{,\rho\rho} - 1 + (\gamma_{ ,\rho})^2 + \
\frac{3}{l^2}\gamma^2\right)  \ .
\label{cons}
\end{equation}

 The  boundary term in the variation of the Hamiltonian (\ref{action})  reads,
\begin{equation}
\left.
-2 \int d\phi d\theta N \gamma_{,\rho\rho} \ \delta
\gamma^2 \ \right|^{r_{++}}_{r_+} \  .
\label{var}
\end{equation}
To arrive at  (\ref{var}) we have taken into account the fact that $\gamma_{,\rho}$
is positive at $r_{+}$ and it is negative at $r_{++}$.

Next, by using the constraint equation,
\begin{equation}
{\cal H}_\perp = 0  \ ,
\label{constraint}
\end{equation}
we rewrite expression  (\ref{var}) as,
\begin{equation}
\left. -2 \int d\phi d\theta N \left(1-\frac{3}{l^2}\gamma^2 -
(\gamma_{,\rho})^2 \right)  \delta \gamma \right|^{r_{++}}_{r_+}  \ ,
\label{var2}
\end{equation}
and  further simplify it, by recalling (\ref{horizon}), to read,
\begin{equation}
 \left. - 2 \int d\phi d\theta  N \  \delta \left( \gamma - \frac{\gamma^3}{l^2} \right)
 \right|^{r_{++}}_{r_+}  \ .
\label{var3}
\end{equation}

Finally, if we set $N=1$ at the boundary, we may write the surface integral
at the boundary as
\begin{equation}
-\delta U \ ,
\label{}
\end{equation}
where the energy $U$ is given by
\begin{equation}
U_+= 8\pi \left( \gamma - \frac{\gamma^3}{l^2} \right) \ ,
\label{up}
\end{equation}
if $r_{++}$ is taken as the boundary, and
\begin{equation}
U_{++}= -8\pi \left( \gamma - \frac{\gamma^3}{l^2} \right) \ ,
\label{upp}
\end{equation}
if the boundary is taken, instead, at $r_+$.

The energy $U$ is conjugated to the Killing time $t$. The choice $N=1$ corresponds to
a particular normalization of the Killing vector, and it is analogous to taking $N^\perp=1$
at spatial infinity in the asymptotically flat case. In this analogy, expressions (\ref{up}), (\ref{upp})
correspond to the ADM surface integral
\begin{equation}
U=\oint \left(g_{ij,i}-g_{ii,j}\right) dS_j \  ,
\label{adm}
\end{equation}
evaluated at spatial infinity in rectangular coordinates.

If one evaluates expressions (\ref{up}),  (\ref{upp})
for the Schwarzschild-de Sitter solution one finds
\begin{equation}
U_+ = m \ ,
\end{equation}
and
\begin{equation}
U_{++} = -m \ ,
\end{equation}
in agreement with the results stated  in the main text.

It would be interesting to develop a discussion along the lines of this appendix
in the presence of rotation. Straightforward efforts to do so did not prove successful.

\end{document}